\documentclass[twoside]{LCWS11}
\usepackage[latin1]{inputenc}
\usepackage[dvips]{graphicx,epsfig,color}
\usepackage{wrapfig,rotating}
\usepackage{amssymb,amsmath,array}
\usepackage{cite}
\begin{document}
\title{
Development of Single- and Double-sided Ladders for the ILD Vertex Detectors} 
\author{J\'erome Baudot$^1$, Olena Bashynska$^{2}$, Nathalie Chon-Sen$^{1}$, Wojciech Dulinski$^{1}$, Franziska Hegner$^{2}$,\\ Marie Gelin-Galibel$^{1}$, Rhorry Gauld$^{3}$, Mathieu Goffe$^{1}$, Joel Goldstein$^{4}$, Ingrid-Maria Gregor$^{2}$,\\ Christine Hu-Guo$^{1}$, Ulrich Koetz$^{2}$, Andrei Nomerotski$^{3}$ and Marc Winter$^{1}$
\vspace{.3cm}\\
1- IPHC/IN2P3/CNRS, Universit\'e de Strasbourg\\
BP 28, F-67037 Strasbourg, France
\vspace{.1cm}\\
2- DESY\\
Notkestr. 85, D-22607 Hamburg, Germany
\vspace{.1cm}\\
3- University of Oxford, Particle Physics\\
Denys Wilkinson Building, Keble Road, Oxford OX1 3RH, UK
\vspace{.1cm}\\
4- University of Bristol\\
Tyndall Avenue, Bristol BS8 1TL, UK
\\
}

\maketitle

\begin{abstract}
We discuss two projects exploring the integration of thin CMOS pixel sensors in order to prototype ladders matching the geometry needed for the ILD vertex detector.\\ 
The PLUME project has designed and fabricated full-size and fully functional double-sided layers which currently reach 0.6~\%~$X_{0}$ and aim for 0.3~\%~$X_{0}$ in mid-2012.\\ 
Another approach, SERNWIETE, consists in wrapping the sensors in a polyimide-based micro-cable to obtain a supportless single-sided ladder with a material budget around 0.15~\%~$X_{0}$. First promising samples have been produced and the full-size prototype is expected in spring 2012. 
\end{abstract}

\section{Introduction}

Two major benefits offered by CMOS pixel sensors (CPS) are their high granularity and low material budget. Indeed, CPS are routinely thinned down to 50~$\mu$m; no other solid-state technology has ever reached such a thickness, encompassing both the sensing and reading elements. In order to integrate CPS into a detector, it is desirable to degrade as little as possible their genuine material budget with the necessary services for the mechanics, electrical powering and data handling.\\
This contribution reviews the progress of two R\&D projects investigating different options for ultra-thin layers with an overall equivalent thickness well below $0.5$~\% of radiation length ($X_{0}$). Both concepts exploit an existing CPS, MIMOSA~26 \cite{refM26}, featuring dimensions ($21.7\times13.9$~mm$^{2}$ surface, 50~$\mu$m thickness) and power dissipation (about 250~mW/cm$^{2}$ in continuous operation) similar to one of the pixel technologies  \cite{refCPSILDproc} proposed for the ILD vertex detector. The modest power dissipated and the possibility to operate the sensor at room temperature, allow to rely on a gentle air flow (a few m/s) for the cooling. Hence, the two technological choices differ mostly through the design of their mechanical support.

\section{Double-sided ladders}

The PLUME (Pixelised Ladder using Ultra-light Material Embedding) collaboration gathers four laboratories (DESY-Hamburg, IPHC-Strasbourg and Universities of Bristol and Oxford) to design, fabricate and evaluate prototypes of double-sided pixelated ladder, matching the geometry required by the inner layer of the vertex detector for the ILD:  a sensitive length of  12.5~cm, a thickness of 2~mm and a material budget around 0.3~\% of $X_{0}$. The ladder design follows a classical approach: six sensors are connected to a low-mass flex-cable to form a module, then two modules are glued on both sides of a mechanical support to form the double-sided ladder. Most of the stiffness of this sandwich-type layer stems from the two modules rather than from the support, which serves essentially as a spacer and can be made of low density material.\\

\begin{figure}[htbp]
\begin{center}
\includegraphics*[width=0.7\columnwidth]{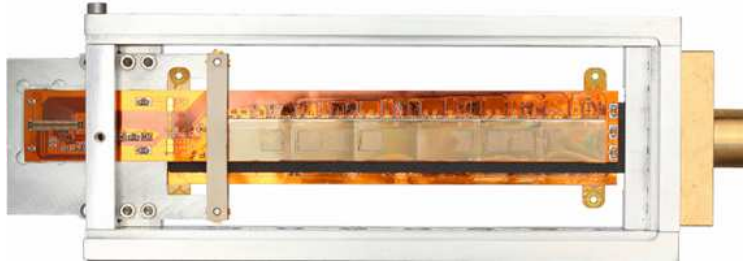}
\caption{Front-view picture of the first full-size PLUME ladder hold in its test support frame. The connector appears on the left side, while the copper piece on the right is the air-inlet for the cooling.}
\label{figPLUME2010Ladder}
\end{center}
\end{figure}

After a very first prototype in 2009 \cite{refPLUMElcws2010}, the first full-scale ladders were designed and fabricated in 2011. The micro-cables are made of two 20~$\mu$m thick metal layers of copper interleaved with 100~$\mu$m thick polyimide. The signal routing focused on electrical safety and resulted in a cable quite wider (24~mm) than the sensor. The spacer material was chosen as silicon carbide foam \cite{refSIC} with an 8~\% density, its width slightly oversized with respect to the sensors for geometrical reasons and to ensure maximal stiffness. The main figures of this ladder, pictured in Figure \ref{figPLUME2010Ladder}, can be summarized as follow: 8~M pixels, mass 10~g equivalent to 0.6~\%~X0 (cross section) and sensitive surface  of $12.7\times1.1$~cm$^{2}$.

Prototypes evaluation started with a mechanical survey, which showed that the surface deviated from a perfectly flat plane with a root mean square around 20~$\mu$m. Secondly, air at ambient temperature and with a flow speed of 3 m/s was blown longitudinally along the ladder length while the 12 sensors were operated. A maximal temperature of 50~$^{\circ}$C on the pixel matrix and 60~$^{\circ}$C on the peripheral circuitry was obtained on the chips.
Though relatively high, this temperature is expected to still guarantee nominal performances. The sensors were successfully operated with threshold values of their integrated discriminators corresponding to respectively 6 and 8 times the typical pixel noise. The rate of fake hits per pixel was measured to sit below $10^{-4}$ and $10^{-5}$ respectively, similarly to observations made with isolated sensors.
In November 2011, the prototype ladder was exposed to the 120~GeV $\pi$ beam of the CERN-SPS at different angles of incidence and all over its surface, to estimate detection efficiency and uniformity as well as spatial resolution. The data analysis is currently under completion.\\


The project next step consists in reducing the overall material budget following two ideas. The silicon carbide spacer density will be decreased to 4~\%, halving this contribution. And a narrower (width 18~mm) flex-cable has been re-designed and re-fabricated in aluminum. These two changes divide the cable contribution to the material budget by a factor respectively two or three, considering either a cross-section or a weighted average. The resulting overall material budget is expected to reach 0.3~\%~$X_{0}$ when a cross section of the ladder is considered and about 0.5~\%~$X_{0}$ when all material are averaged with respect to the sensitive width of 10~mm.\\
In order to predict the impact of this new design on the mechanical and thermal behavior of the ladder, a finite element analysis was conducted on a model of the first PLUME ladder including the air flow cooling. The average temperatures over the surface of the sensors reproduced our measurements satisfactorily. Regarding the mechanical stiffness, the predicted frequency of the ladder first vibration mode was 250~Hz. The simulation of the new ladder did not change those figures and hence validated our design.\\
First aluminum micro-cables have been produced at the CERN workshop early in 2012 and are under evaluation. The new ladder prototype will be evaluated under beam conditions in November 2012.

\section{Sensor embedding}

\begin{figure}[htbp]
\begin{center}
\includegraphics*[width=0.75\columnwidth]{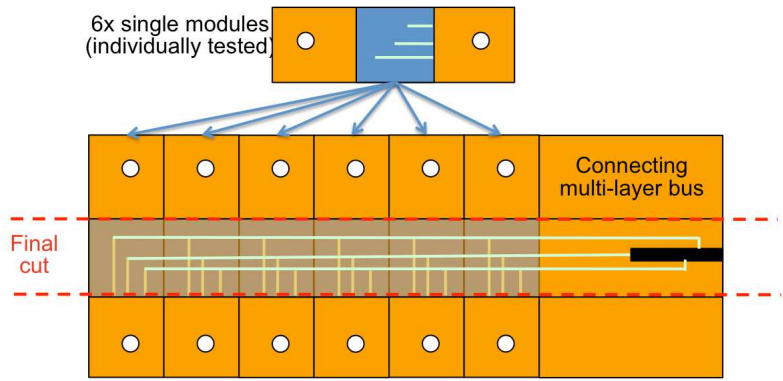}
\caption{Sketch of the assembly of a 6-sensor single-sided ladder from the former embedding of single sensors into a micro-cable.}
\label{figCERNWIETEsketch}
\end{center}
\end{figure}

The particularly low thickness of CPS allows to consider their embedding directly inside the multi-layer micro-cable, at the fabrication phase of the latter. The process consists in gluing the chips on the first polyimide substrate layer prior to proceed with the usual assembly of the subsequent layers of metal and insulator used for such cables. The metal deposition step directly connects the chip pads to metal traces exactly as vias connect two metal layers on such cables and avoids space-consuming wire-bonding. A major benefit of this process is the lessening of the mechanical stress on the sensors, which is taken away by the polymer films wrapping it. The final object is a supportless single-sided ladder which can potentially adapt to non-planar geometries.\\
However, a major practical difficulty in the embedding of several sensors in a cable resides in the alignment of their respective pad rows with the metal traces. The CERN printed circuits workshop undertook solving this problem by wrapping one sensor at a time in a single metal layer cable offering new redistributed connection pads. Then several such single-sensor cables can be connected altogether with a new long cable with additional metal layers (two to three). The process flow of this project, named SERNWIETE (SEnsor Row Neatly Wrapped In an Extra-Thin Envelope), is illustrated in Figure \ref{figCERNWIETEsketch}.

First prototypes of single-sensor cables were produced late in 2011 and early in 2012. They were powered on with the mean of a probe-station to check basic electrical functionalities. The first multi-sensor cable is expected for Spring 2012. The current design relies on 50~$\mu$m thick MIMOSA~26 sensors, 4 layers of aluminum and 20 to 50~$\mu$m polyimide films. The material budget of this first supportless module reaches 0.2~\% of $X_{0}$. Further optimization is expected to result in about 0.15~\%~$X_{0}$ of total material budget.

\section{Summary}

We reported the fabrication by the PLUME collaboration of a first fully functional double-sided ladder. This prototype matches the geometrical requirements of the ILD vertex detector and its current performances augur well for the design and fabrication in 2012 of a double-sided ladder featuring the desired material budget, ${\cal O}(0.3)$~\%~$X_{0}$.\\ 
This first exercise based on an existing CPS allowed to setup the necessary infrastructures among the participating laboratories, to design, simulate, build and evaluate new prototypes equipped with either the same or alternative sensors.\\
The SERNWIETE project exploits further the new integration possibilities opened up by the CMOS pixel sensor technology to reach smaller material budget per layer (below 0.2~\%~$X_{0}$) and cover non-planar surfaces.\\
While the SERNWIETE project has not yet delivered its first full prototype, expected for 2012, the PLUME prototypes represent an important milestone for the ILD vertex detector technical development route. Among others, six to eight PLUME double-sided ladders will be operated over long periods in the coming years within the framework EU-FP7 AIDA \cite{refAIDA} project.

\section{Acknowledgments}

We thank Ryan Page, from the University of Bristol, for having provided the essential expertise to conduct the mechanical survey.

\section{Bibliography}


\begin{footnotesize}

\end{footnotesize}


\end{document}